\documentclass{iaus}
\usepackage{graphicx}
\usepackage{graphicx,bm,url}
\usepackage{natbib}
\newcommand{\EQ}{\begin{equation}}
\newcommand{\EN}{\end{equation}}
\newcommand{\EQA}{\begin{eqnarray}}
\newcommand{\ENA}{\end{eqnarray}}

\newcommand{\Fig}[1]{Figure~\ref{#1}}

{}
{}
{}

{}
{}
{}
{}
{}
{}
{}
{}
{}
{}
{}
{}
{}
{}
{}
{}
{}

{}
{}
{}

{}

{}
{}

\newcommand{\xx}{\bm{x}}

\newcommand{\uu}{\mbox{\boldmath $u$} {}}

\newcommand{\nab}{\mbox{\boldmath $\nabla$} {}}
\newcommand{\OO}{\bm{\Omega}}

\def\csz{c_{\rm s0}}

\newcommand{\yaraa}[3]{ #1, \textit{ARA\&A,} \textit{#2}, #3}

\newcommand{\ymn}[3]{ #1, \textit{MNRAS,} \textit{#2}, #3}

\newcommand{\pproc}[4]{ #1, in \textit{#2}, ed.\ #3 (#4), (in press)}

\title[Vorticity from irrotationally forced flow] %% give here short title %%
{Vorticity from irrotationally forced flow}

\author[Fabio Del Sordo \& Axel Brandenburg]   %% give here short author list %%
{ Fabio Del Sordo
 and Axel Brandenburg}

\affiliation{NORDITA,
Roslagstullsbacken 23, SE-10691 Stockholm, Sweden; and \\
Department of Astronomy, Stockholm University,
SE 10691 Stockholm, Sweden}

\pubyear{2010}
\volume{271}  %% insert here IAU Symposium No.
\pagerange{119--126}
\jname{Astrophysical Dynamics: From Stars to Galaxies}
\editors{A.C. Editor, B.D. Editor \& C.E. Editor, eds.}
\begin{document}

\maketitle

\begin{abstract}
In the interstellar medium the turbulence is believed to be forced mostly through supernova explosions.
In a first approximation these flows can be written as a gradient of a potential being thus devoid of vorticity.
There are several mechanisms that could lead to vorticity generation, like viscosity and baroclinic terms, 
rotation, shear and magnetic fields, but it is not clear how effective they are, neither is it clear whether
the vorticity is essential in determining the turbulent diffusion acting in the ISM.
Here we present a study of the role of rotation, shear and baroclinicity in the generation of vorticity in the ISM.

\end{abstract}

\begin{figure}[t!]
\begin{center}
 \includegraphics[width=\linewidth]{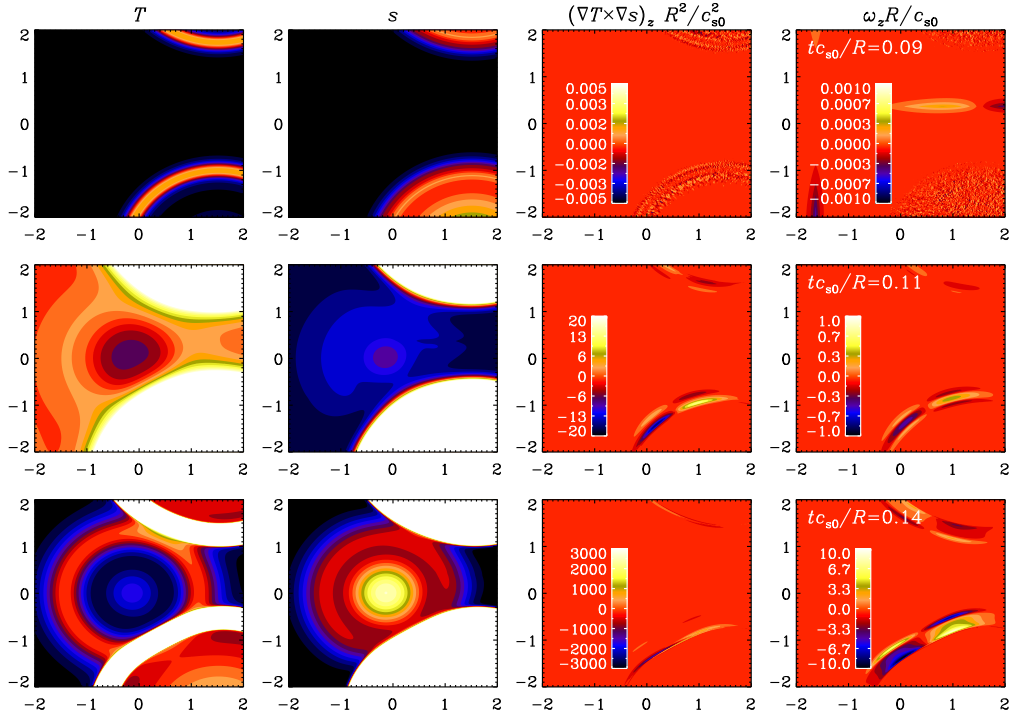}\qquad
 \caption{Images of temperature, entropy, baroclinic term $(\nab T\times\nab s)_z$ and
normalized vertical vorticity for a two-dimensional
run with $\delta t_{\rm force}\csz/R=0.1$ at an instant shortly before
the second expansion wave is launched (top row), and shortly after the
second expansion wave is launched (second and third row).
In the second and third row the vorticity production from the baroclinic term is clear,
while in the top row, $(\nab T\times\nab s)_z$ and $\omega_z$
are just at the noise level of the calculation.
Shock surfaces are well localized and
the zones of maximum production of vorticity are those in which the fronts
encounter each other.
Adapted from \cite{DSB10}.
}\label{p2d}
\end{center}
\end{figure}

The study of the interstellar medium (ISM) is strictly connected with that of 
supernovae explosions, being this one of the most important phenomena in 
determining its dynamic. Such events act on scales up to  $\sim 100$ pc injecting 
enough energy to sustain turbulence with velocities of $\sim 10$ km/s.
An understanding of the generation of turbulence and vorticity in the ISM is a basic requirement 
in order to formulate a theory for the production of interstellar magnetic fields.
The first step to perform numerical simulations of these explosion is
to assume the presence of pure potential forcing. Indeed a supernova explosion 
can be seen at first glance as a pure spherical expansion that is driving turbulence in the ISM.
In our simulations we use the \textsc{Pencil Code},
http://pencil-code.googlecode.com/
We simulate spherical expansions following the work of \cite{MB06}. 
We analyze flows that are only weakly supersonic and use a constant and 
uniform viscosity in an unstratified medium.
We solve the Navier-Stokes equations in the presence of viscosity
and with a potential forcing $\nabla\phi$ where $\phi$ is given by
\EQ
  \phi(\xx,t)=\phi_0\ N\exp\left\{[\xx-\xx_{\rm f}(t)]^2/R^2\right\}.
\EN
Here $\xx=(x,y,z)$ is the position vector,
$\xx_{\rm f}(t)$ is the random forcing position,
$R$ is the radius of the Gaussian, and $N$ is a normalization factor.
We consider two forms for the time dependence of $\xx_{\rm f}$.
First, we take $\xx_{\rm f}$ such that the forcing is
$\delta$-correlated in time. Second, we include a forcing time
${\delta{}t}_{\rm force}$ that defines the interval during which $\xx_{\rm f}$
remains constant.
We study then the effect of rotation on such environment.
Rotation causes the action of the Coriolis force, $2\OO\times\uu$,
in the evolution equation for the velocity.
We investigate flows with Reynolds numbers up to 150. We find that there is a clear production 
of small scale vorticity when we use both a finite and $\delta$-correlated forcing. In this last case
we observe some spurious vorticity also at very low Coriolis number, due probably to some numerical artifact.
We then consider the effects of shear on the potential flow.
In the presence of linear shear with $\uu^S=(0,Sx,0)$, the
evolution equation for the departure from the mean shear attains
additional terms, $-\uu^S\nab\cdot\uu-\uu\cdot\nab\uu^S$.
Also in this case we observe the production of vorticity but we do not
find clear results for small values of the shear. 
We have thus observed a production of vorticity due to rotation and shear, but
these effects are not able to produce enough vorticity
under the physical condition of the ISM, like those described by \cite{Beck96}. 
When we relax the isothermal condition the system is then also under the
action of a baroclinic term. This is  proportional
to the cross product of the gradients of pressure and density
and emerges when taking the curl of the pressure
gradient term, $\rho^{-1}\nab p$.
The results we obtain are shown in \Fig{p2d}.

It turns then out that the baroclinic term is more efficient in the production of the vorticity.
Moreover the biggest amount of vorticity is observed when shock fronts encounter each other.

Speaking about possible dynamo action, as pointed out by \cite{BD09}, the presence of vorticity 
does not seem to affect the diffusion of magnetic fields differently than a complete irrotational turbulence.
Nevertheless the vorticity plays an important role in dynamo processes so it is still important
to address the problem of the generation of the vorticity investigating the possible role of other effects.

\end{document}